\documentclass[prb,aps,twocolumn]{revtex4}

\usepackage{graphicx}
\usepackage{dcolumn}
\usepackage{bm}

\begin{document}

\title{Persistent superfluid phase in a three-dimensional quantum XY model with ring exchange}

\author{R.~G.~Melko}
\affiliation{Department of Physics, University of California Santa
Barbara,
California 93106}

\author{D.~J.~Scalapino}
\affiliation{Department of Physics, University of California Santa
Barbara,
California 93106}

\date{\today}

\begin{abstract}

We present quantum Monte Carlo simulation results on a quantum $S=1/2$ XY model 
with ring exchange (the $J$-$K$ model) on a three-dimensional simple cubic 
lattice.  We first characterize the ground state properties of the pure 
XY model, obtaining estimations for the energy, spin stiffness and 
spin susceptibility at $T=0$ in the superfluid phase.  With the ring exchange, 
we then present simulation data on small lattices which suggests that the superfluid 
phase persists to very large values of the ring exchange $K$, without 
signatures of a phase transition. We comment on the consequences 
of this result for the search for various exotic phases in three dimensions.

\end{abstract}

\maketitle

The search for microscopic spin 1/2 Hamiltonians that exhibit a spin liquid 
phase continues with the hope that these studies will provide insight into
the type of materials that can support such a phase.  Focus has recently
turned to multi-particle ring exchange models,\cite{SSSD02,Ring1,ALMebl}  
which have been identified as
promising candidates for the presence of spin liquid as well as other exotic 
phases and critical points.
One such Hamiltonian is an easy-plane $S=1/2$ $J$-$K$ model on the square 
lattice, with an XY exchange coupling $J$ and a four-site ring exchange 
coupling $K$.  Quantum Monte Carlo studies\cite{SSSD02}  of this model 
have shown that, 
at a critical value $(K/J)_c$, there appears to be a continuous quantum critical
point separating an XY-superfluid phase from a valence bond solid (VBS) phase.
Recently, Senthil {\it et al.}\cite{DQCP} have proposed that at this type of 
critical point -- separating conventional phases described in terms of different 
broken 
symmetries -- one can have an isolated spin-liquid point characterized by an 
emergent global $U(1)$ symmetry and deconfined spinon excitations.  
In two dimensions, the theory finds that the $U(1)$ spin liquid is unstable 
away from the critical point because of the proliferation of monopole 
excitations, which are only absent precisely at $(K/J)_c$.
They suggest\cite{DQCP} that it is this instability
that is responsible for the formation of the VBS phase seen in the
simulations\cite{SSSD02} for $K/J > (K/J)_c$.
However, in three dimensions, monopole excitations can be gapped, 
so within the theoretical framework\cite{DQCP} an 
extended region of stable $U(1)$ spin liquid phase could exist.\cite{3DU1}
Motivated by this possibility, we have carried out a quantum Monte Carlo study of the
simple cubic $J$-$K$ model to look for signatures of a zero-temperature phase 
transition out of the superfluid phase (at large $J/K$) to a quantum disordered 
or insulating phase (at large $K/J$).

\begin{figure}[ht]
\includegraphics[height=2.3cm]{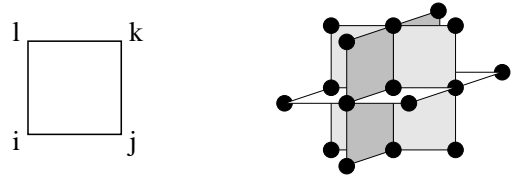}
\caption{
The labeling of the plaquette indices in Eq.~(\ref{ham}) are illustrated on 
left.
At right, three intersecting plane segments of the cubic lattice show that each
spin site shares six bonds and twelve plaquettes with other
sites.
}
\label{plaq}
\end{figure}

The Hamiltonian under study is
\begin{equation}
H = -J\sum\limits_{\langle ij\rangle} B_{ij}
    -K\sum\limits_{\langle ijkl\rangle} P_{ijkl},
\label{ham}
\end{equation}
where  bond and plaquette exchange operators are defined as
\begin{eqnarray}
B_{ij} & = & S^+_iS^-_j + S^-_iS^+_j = 2(S^x_iS^x_j + S^y_iS^y_j), 
\label{bond} \\
P_{ijkl} & = & S^+_iS^-_jS^+_kS^-_l + S^-_iS^+_jS^-_kS^+_l.
\end{eqnarray}
Here, $\langle ij\rangle$ denotes a pair of nearest-neighbor sites and
$\langle ijkl\rangle$ are sites on the corners of a plaquette 
(see Fig.~\ref{plaq}).   
In three dimensions, each spin is shared by six bonds and twelve plaquettes, 
which form the faces of the simple cubic lattice.
For $K=0$ this is the standard quantum spin-1/2 $XY$-model, or, 
equivalently, a hard-core boson model at half-filling, studied
previously using quantum Monte Carlo techniques by Pedersen and 
Schneider.\cite{PSmc3D}
The model is analogous to a quantum lattice gas model of He$^4$, and undergoes a
lambda-transition at $T/J \approx 2$ to a bulk superfluid 
state.\cite{PSmc3D,Betts}
The $K$-term corresponds to retaining only the $x$- and $y$-spin 
operators of the full four-spin cyclic exchange studied in relation to some 
high-$T_c$ parent compounds.\cite{CupRing}

Our simulations use the stochastic series expansion (SSE) quantum Monte Carlo 
method,\cite{SK91,AWS56} which is based on importance sampling of elements of 
the power series
expansion of the partition function.  The method is numerically exact within
statistical error bars, and has broad applicability to a wide range of spin 
and boson models where the sign problem can be avoided.  
In order to efficiently sample off-diagonal $J$ terms in the Hamiltonian, the 
recently developed ``directed-loop''\cite{San99,SS02} algorithm is used.  However, for large 
values of the parameter $K$, the loop moves become ineffective, and a new 
sampling procedure was developed.\cite{AWSmbc}  This new 
``multi-branch cluster'' update is a Swendson-Wang or Wolff-type sampling scheme
generalized to flip clusters of plaquette operators in the SSE operator-list. 
It is observed to significantly decrease autocorrelation times of the 
simulations for large $K$, even for $K/J \rightarrow \infty$ in two dimensions.  
Details of the SSE algorithm, simulation procedure and calculation of physical
observables can be found in Ref.~\onlinecite{AWSmbc}.

The energy per spin is simply calculated in terms of the SSE estimator for the order of the power series expansion, $n$, and the temperature:
\begin{equation}
E=-\frac{\langle{n}\rangle T}{N},
\label{nrg}
\end{equation}
where $N=L^3$ is the system size.
The main quantity of interest in this paper is the spin stiffness,
\begin{eqnarray}
\label{d2E}
\rho_s&=&\frac{\partial^2 E(\phi)}{\partial \phi^2} \\ 
&=& \frac{T \left({ \langle W^2_x \rangle + \langle W^2_y \rangle + \langle W^2_z \rangle }\right)}{3N},
\label{ss}
\end{eqnarray}
where $\phi$ is a twist imposed on bonds in either the $x$, $y$ or $z$
lattice directions.
In Eq.~(\ref{ss}) we calculate the derivative at $\phi=0$ using
the winding number estimators $W_{x,y,z}$ which measure the net spin current across the periodic boundaries in each direction.\cite{AWS56,AWSmbc}
We have also calculated in some cases the spin susceptibility $\chi$, easily 
obtainable in the $S^z$ basis using the squared magnetization 
estimator,
\begin{equation}
\chi = \frac{1}{T N} \left< { \left( {\sum_{i=1}^N S_i^z} \right) ^2 }\right>.
\label{chi}
\end{equation}

In three dimensions, even the smallest useful lattice sizes 
have Hilbert spaces far too large to allow an exact diagonalization study
to be performed.
In order to test the Monte Carlo and to
characterize the $K=0$ ground state, we study the asymptotics of the pure
XY model, setting $J=0.5$ in the Hamiltonian Eq.~(\ref{ham}).
Second-order spin-wave calculations carried out by Weihong 
{\it et al.}\cite{WOH} give numerical values for the ground state energy per 
spin and spin susceptibility as $E_0 \approx -0.7914$ and
$\chi \approx 0.1517$, without error bars. Although we are not aware of any 
detailed numerical results for the zero-temperature spin stiffness, mean field 
theory predicts a value of $\rho_s = S^2=0.25$ using 
Eq.~(\ref{d2E}).  One might expect that quantum fluctuations slightly increase
the value of $\rho_s$ in the full $S=1/2$ XY model, as observed in two 
dimensions.\cite{AH2dXY}

\begin{figure}
\includegraphics[height=6cm]{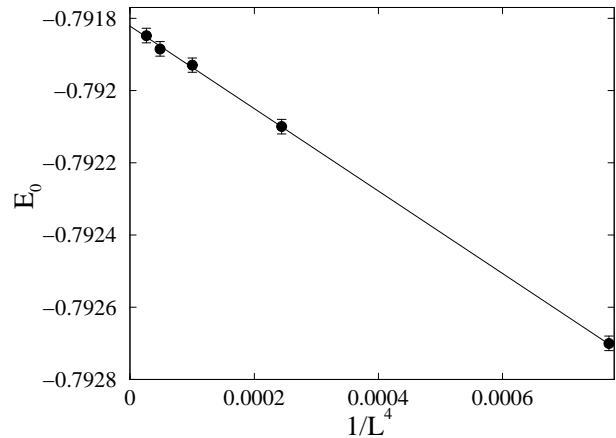}
\caption{
Ground-state energy versus $1/L^4$ for system sizes $L=6$,8,10,12 and 14.
The curve is a fit to Eq.~(\ref{Efss}).
}
\label{Escale}
\end{figure}
We simulated the pure XY model ($K=0$, $J=0.5$) for several lattice sizes, 
L$=6,8,10,12$, and 14, using the full SSE code, for several values of 
$T /J$ down to 0.04.  
The relevant estimators either showed absolute 
temperature convergence to within error bars, or were extrapolated to $T=0$.  
Data obtained from the SSE simulation for the ground-state energy per spin, 
$E_0$, is most accurate.  Chiral perturbation theory\cite{HNfss}
predicts that the energy depends on system size and
spatial dimension ($d$) as
\begin{equation}
E_0(L) -E_0 \sim \frac{C}{L^{d+1}}, 
\label{Efss}
\end{equation}
where $C$ is some constant.
This finite-size dependence has been shown to 
be in good agreement with similar quantum Monte Carlo studies of the 
two-dimensional (2D) XY model.\cite{AH2dXY}
Setting $d=3$ in Eq.~(\ref{Efss}), we use our data to
obtain an extrapolation to $L \rightarrow \infty$, giving $E_0=-0.79182(2)$.
The fit is illustrated in Fig.~\ref{Escale}, which has a minimum in the 
chi-squared value per degree of freedom
for $C=-1.14$.  
Data for the spin susceptibility and spin stiffness have considerably larger
error bars, within which the larger system size data is converged,
making a finite-size scaling extrapolation irrelevant.
Our best estimates for the ground state values in these cases are $\chi = 0.1500(2)$
and $\rho_s = 0.2623(2)$

\begin{figure}[ht]
\includegraphics[height=6.2cm]{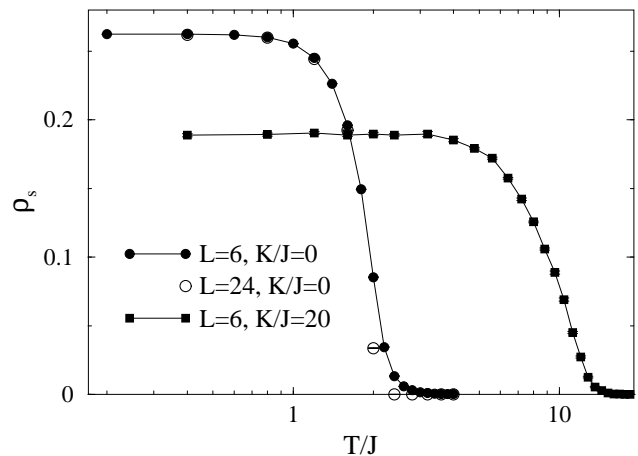}
\caption{The finite-temperature behavior of the spin stiffness.  
Data for both $L=6$ and $L=24$ is included for $K/J=0$ to illustrate 
finite-size effects at the lambda-transition. 
}
\label{L6rho}
\end{figure}

\begin{figure}[h]
\includegraphics[height=6cm]{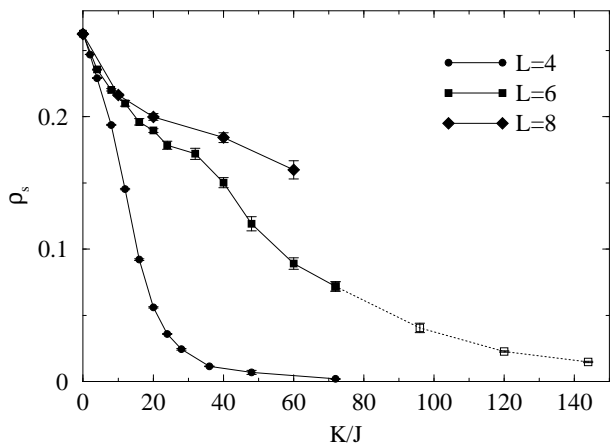}
\caption{Ground-state spin stiffness for the 3D $J$-$K$ model. 
The XY exchange coupling was set to a constant, $J=0.5$.
Open symbols indicate data which showed signs of ergodicity loss, and therefore
error bars may be under-estimated.
}
\label{468rho}
\end{figure}

Turning to the full model, we performed simulations including the $K$ term in Eq.~(\ref{ham}), keeping $J$ fixed at $1/2$.  
First, the $T \rightarrow 0$ temperature convergence was established for each value of $K/J$ simulated.  Fig.~\ref{L6rho} compares the finite-temperature behavior of the spin stiffness for an $L=6$ system at $K/J=20$ versus the XY-point at $K/J=0$.  
Results for the ground-state spin stiffness versus $K/J$ are illustrated in Fig.~\ref{468rho}.  Data for $L=4$ and $L=6$ up to $K/J=48$ was taken at temperatures below $T/J = 0.5$, depending on the specific $K/J$ where good $T \rightarrow 0$ convergence was established.  However, the three largest $K/J$ values for $L=6$ began to show sticking, evidence that the Monte Carlo time-scales are getting very large or ergodicity is being lost.  To combat this, data for $L=8$ was taken at slightly higher temperature, $T/J < 2.0$ (for large $K/J$), 
which was nonetheless within the range of the zero-temperature convergence.

For the 2D $J$-$K$ model, the zero-temperature critical point separating the superfluid phase from the VBS occurs at $(K/J)_c \approx 8$.\cite{SSSD02}  However, as illustrated in Fig.~\ref{468rho}, for the three-dimensional (3D) case the superfluid phase is clearly much more robust.
In particular, extensive simulations for the $L=6$ system, out to $K/J=144$ fail to show an abrupt transition out of the superfluid phase.  Similarly, data for $L=8$ remains strongly superfluid for $K/J$ up to at least 60.  
As evident from Fig.~\ref{468rho}, finite-size effects for the spin stiffness
become much more pronounced as $K/J$ gets large.
Further, observation of the staggered magnetization showed no indications of
the onset of any N\'eel (or boson charge-density wave) ordered state at any $K/J$
which was simulated.
Unfortunately, algorithm performance becomes less efficient for this model for large values of $K/J$.  This, coupled with the fact that the lattice-plaquette coordination contributes an extra factor of three in the CPU time size scaling of the algorithm in three dimensions as compared to two dimension, makes a more detailed study of larger lattice sizes impractical at this time.

The consistency of data for the two largest system sizes
points toward robust superfluidity in the system well above the value for 
$(K/J)_c$ where the onset of insulating behavior occurs
in the 2D model.\cite{SSSD02}
Although impossible to prove without a detailed finite-size scaling study, we believe that the data  supports the conjecture that the superfluid ground state
persists in the model, becoming critical only at $K/J \rightarrow \infty$. 
This of course leaves the interesting question regarding the nature of the $K/J \rightarrow \infty$ phase open for future study.  One suggestion is that the 
$J=0$ model actually has long-range in-plane ferromagnetic order, 
in accord with a simple spin-wave analysis.\cite{LBpc}  The critical value for 
the spin stiffness in this case would be $\rho_s = 0$, however for finite $J$ as
in Fig.~\ref{468rho}, $\rho_s$ may be non-zero in the thermodynamic limit.

In summary, we have presented data from SSE quantum Monte Carlo simulations of 
the 3D simple cubic $J$-$K$ model.  Simulations of the XY point
reveal a ground state energy, spin susceptibility, and spin stiffness
of
\begin{eqnarray}
E_0 &=&-0.79182(2), \nonumber \\
\chi &=& 0.1500(2), \nonumber \\
\rho_s &=& 0.2623(2), \nonumber
\end{eqnarray}
consistent with predictions.  In particular, the ground state energy
agrees with second-order spin-wave theory\cite{WOH} to within $0.05\%$.  
The full $J$-$K$
model appears to remain superfluid for all finite $K/J$, with no phase 
transition signature up to $K/J = 144$.  The large $K$ insulating phase is still
unknown, however no indications of either a charge-density wave or any quantum
disordered state are present.  
The robustness of the superfluid phase is consistent with the notion that the 
relative strength of quantum fluctuations introduced by the ring-exchange ($K$) term is reduced as one increases the dimensionality of the system.
Finally, the failure of this model to support an emergent 
$U(1)$ spin liquid illustrates the difficulty in
realizing exotic phases in microscopic quantum models suggested by
arguments arising from generalized actions.

\acknowledgments{
The authors would like to thank T.~Senthil for motivating discussions.  We appreciate extensive and helpful collaboration with A.~Sandvik.  Thanks also to 
L.~Balents, A.~Burkov, M.~P.~A.~Fisher, M.~Hermele, and O.~Motrunich
for useful conversations.
Supercomputer time was provided by NCSA under grant number DMR020029N.  
This work was also
supported by the Department of Energy, Grant No. DOE85-45197.

\end{document}